\documentclass[aps,prl,notitlepage,superscriptaddress,reprint]{revtex4-1}
\usepackage[utf8]{inputenc}
\usepackage{siunitx}

\usepackage{ulem}
\usepackage{amsmath}

\newcommand{\qav}[1]{\langle {#1} \rangle}
\newcommand{\ket}[1]{\left \lvert {#1} \right \rangle}

\begin{document}

\title{Partitioning of on-demand electron pairs}

\author{Niels Ubbelohde}
\affiliation{Institut f\"{u}r Festk\"{o}rperphysik, Leibniz Universit\"{a}t Hannover, 30167 Hannover, Germany}

\author{Frank Hohls}
\affiliation{Physikalisch-Technische Bundesanstalt, 38116 Braunschweig, Germany}

\author{Vyacheslavs Kashcheyevs}
\affiliation{Faculty of Physics and Mathematics, University of Latvia, LV-1002, Riga, Latvia}

\author{Timo Wagner}
\affiliation{Institut f\"{u}r Festk\"{o}rperphysik, Leibniz Universit\"{a}t Hannover, 30167 Hannover, Germany}

\author{Lukas Fricke}
\affiliation{Physikalisch-Technische Bundesanstalt, 38116 Braunschweig, Germany}

\author{Bernd K\"{a}stner}
\affiliation{Physikalisch-Technische Bundesanstalt, 38116 Braunschweig, Germany}

\author{Klaus Pierz}
\affiliation{Physikalisch-Technische Bundesanstalt, 38116 Braunschweig, Germany}

\author{Hans W. Schumacher}
\affiliation{Physikalisch-Technische Bundesanstalt, 38116 Braunschweig, Germany}

\author{Rolf J. Haug}
\affiliation{Institut f\"{u}r Festk\"{o}rperphysik, Leibniz Universit\"{a}t Hannover, 30167 Hannover, Germany}

\date{\today}

\begin{abstract}

We demonstrate the high fidelity splitting of electron pairs emitted on demand from a dynamic quantum dot by an electronic beam splitter. The fidelity of pair splitting is inferred from the coincidence of arrival in two detector paths probed by a measurement of the partitioning noise.
The emission characteristic of the on-demand electron source is tunable from electrons being partitioned equally and independently to electron pairs being split with a fidelity of 90~\%. For low beam splitter transmittance we further find evidence of pair bunching violating statistical expectations for independent fermions.

\end{abstract}

\maketitle

The on-demand generation and separation of entangled photon pairs are key components for quantum information processing in quantum optics \cite{Kim1999, Stevenson2006, Lang2013}. In an electronic analog \cite{Yamamoto2012} the decomposition of electron pairs in scalable solid state systems represents an essential building block for employing the quantum state of ballistic electrons in electron quantum optics experiments \cite{Ji2003, Neder2007, Altimiras2009}. 
While the scattering of electrons in a mesoscopic beam splitter has been used to probe the particle statistics of stochastic sources in fermion Hanbury Brown and Twiss (HBT) experiments \cite{Henny1999, Oliver1999}, the recent advent of on-demand sources \cite{Feve2007, Bocquillon2012} further offers the possibility to synchronize the emission \cite{Fletcher2013} in order to achieve indistinguishability between multiple sources as evidenced by quantum collisions \cite{Liu1998} in Hong-Ou-Mandel experiments \cite{Bocquillon2013}. Cooper pairs impinging stochastically at a mesoscopic beam splitter have been successfully partitioned \cite{Wei2010, Herrmann2010, Hofstetter2009, Hofstetter2011, Das2012, Schindele2012} verified by measuring the coincidence of arrival at two detectors \cite{Wei2010, Herrmann2010, Das2012, Schindele2012}. Here we demonstrate the splitting of electron pairs generated on demand \cite{Fletcher2013} from a dynamic quantum dot. The coincidence correlation measurements in a HBT configuration combined with a highly regular incident beam allow the reconstruction of the full counting statistics of the electron pair transmission, which reveals regimes of statistically independent, distinguishable or correlated partitioning and has been envisioned as a source of information on the quantum state of the electron pair \cite{Burkard2000, Samuelsson2006, Hassler2007, Hassler2008, Wahl2014}. The high achievable pair-splitting fidelity of up to \SI{90}{\%} opens the path for a future on-demand generation of spin-entangled electron pairs from a suitably prepared two-electron quantum dot ground state.

\begin{figure}
\includegraphics{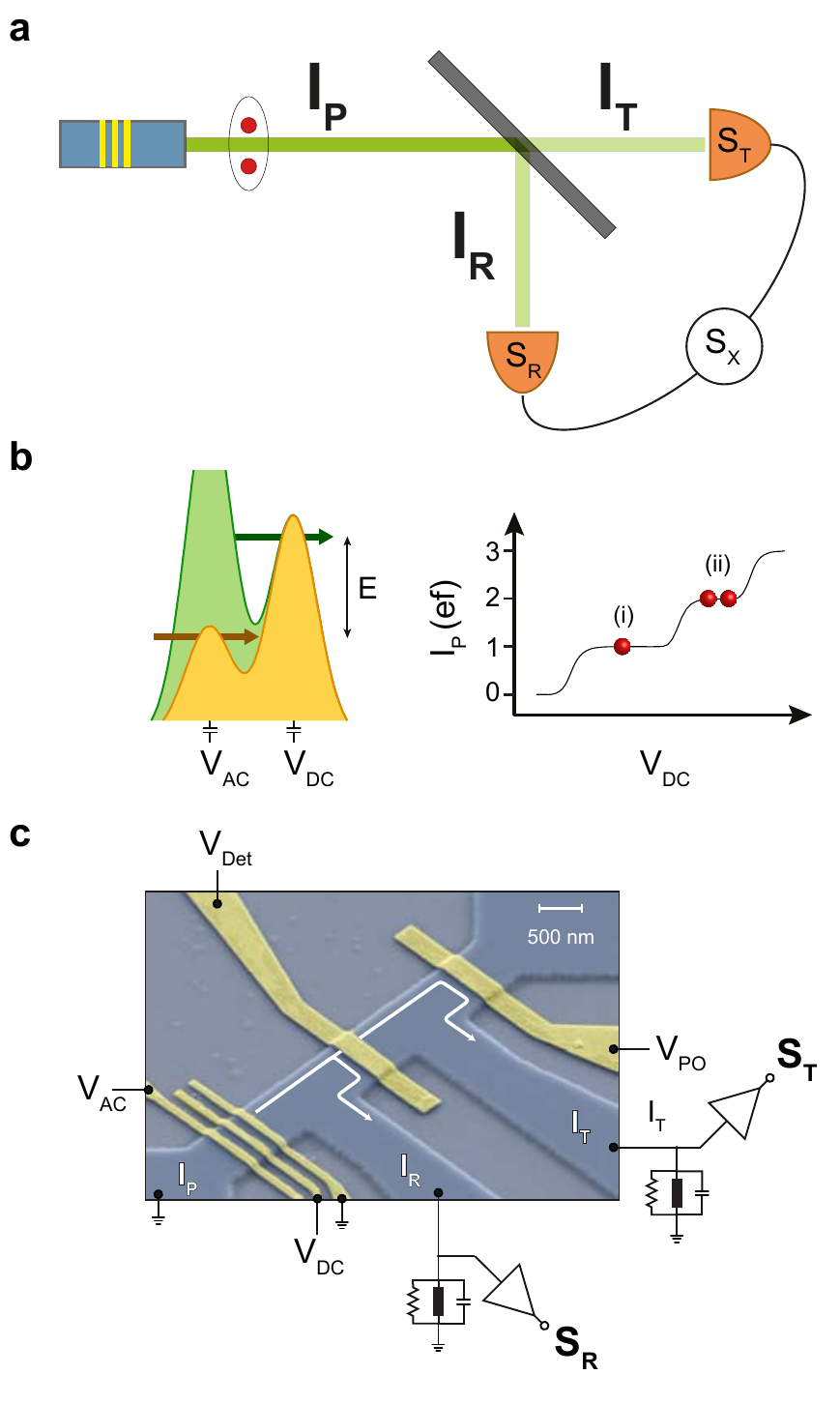}
\caption{\label{fig:fig1}
%\textbf{Measurement setup}
\textbf{a} Schematic diagram of the experimental setup following the Hanbury Brown and Twiss geometry. The current $\mathrm{I_P}$ of the triggered electron source is split into a reflected part $\mathrm{I_R}$ and a transmitted part $\mathrm{I_T}$. In order to measure the coincidence of arrival the cross-correlation between the time-dependent fluctuations $\mathrm{I_R(t)}$ and $\mathrm{I_T(t)}$ are detected giving rise to the cross-correlation noise power $\mathrm{S_X}$ and the single detector channel noise power $\mathrm{S_R}$ and $\mathrm{S_T}$. \textbf{b} Energy diagram of the loading (orange) and unloading (green) phase of the dynamic quantum dot resulting in a quantized current $\mathrm{I_P}$. \textbf{c} Typical current characteristic showing plateaus corresponding to the emission of single electrons (i) and electron pairs (ii) of the driving waveform applied to $\mathrm{V_{AC}}$. \textbf{d} Micrograph of the sample geometry. The gates controlled with the voltages $\mathrm{V_{AC }}$ and $\mathrm{V_{DC}}$ form the electron source. White lines indicate chiral edge-channels, which guide the electrons to the detector barrier defined by $\mathrm{V_{Det}}$. The channel exit is pinched-off ($\mathrm{V_{PO}}$) ensuring that all transmitted electrons are recorded by $\mathrm{I_T}$.}
\end{figure}

The few-electron source is based on a single parameter non-adiabatic quantized charge pump \cite{Blumenthal2007, Kaestner2008, Giblin2012}, which enables the deterministic generation of single electrons and electron pairs with tunable emission energy \cite{Leicht2011, Fletcher2013}.
The non-equilibrium electrons propagate along the edge of a quantum Hall sample with minimal inelastic scattering. The device and the measurement setup are sketched in Fig.~\ref{fig:fig1}. An energy-selective detector barrier splits the incoming beam of electrons into two detector paths.  The coincidence of arrival of electrons in the two detector channels leads to positive correlation between the time-dependent current signals. These correlations are inferred from a measurement of the zero-frequency cross-correlation shot noise. While an oscillator controlled electron source is noiseless \cite{Maire2008}, the splitting of electron pairs generates partitioning noise and enables a tomography of the probability distribution for the partitioning outcomes within each emission cycle.

%\section{Methods}

The electrons are sourced by a dynamic quantum dot formed by two metallic gates over an \SI{900}{\nano m} wide etched channel of a two-dimensional electron gas (2DEG) \SI{90}{nm} beneath the surface of a GaAs/AlGaAs heterostructure \cite{Kaestner2008}. The carrier density and mobility of the 2DEG are \SI{1.6e6}{cm^2/Vs} and \SI{2.6e11}{cm^{-2}} respectively. Controlled by the modulation of the entrance gate a defined number of electrons are loaded into the quantum dot and ejected over the exit barrier into the channel. The entrance gate voltage is driven by an arbitrary waveform generator enabling the triggered emission of electrons. The exit gate is held at a fixed potential. Due to an applied perpendicular magnetic field of \SI{12}{T} the emitted electrons follow the sample edge to a detector barrier, where they are transmitted or reflected into two detector channels depending on their emission energy in relation to the energetic barrier height. The value for the magnetic field was chosen in order to suppress the energy relaxation of the electrons on the \SI{2}{\micro m} long path towards the detector barrier. For lower values of magnetic field inelastic scattering sets in, dominated by longitudinal optical phonon coupling \cite{Taubert2011, Sivan1989} and visible as steps in the transmitted current with a period of \SI{36}{meV}.

A measurement of the transmitted current $\mathrm{I_T}$ allows to determine the average fraction of reflected electrons. In addition, the fluctuations of the transmitted and reflected current are recorded. The real part of the cross-correlation of the two detector signals is integrated over 20 minutes and averaged within a frequency window between \SI{500}{kHz} and \SI{3}{MHz} yielding the zero-frequency shot noise power $\mathrm{S_X}$, which is determined by the partitioning noise.

The energetic height of the detection barrier is calibrated by driving a constant current across the exit barrier. With the entrance gate held at ground potential the energy of the electrons emitted into the channel is then controlled by the exit gate and approximately given by the bias voltage $\mathrm{V_{Bias}}$ across the exit gate. The combined measurement of $\mathrm{I_T}$ and $\mathrm{V_{Bias}}$ as a function of the voltages applied to the exit gate and the detector gate relates the transmission energy of the electrons to the detection gate voltage $\mathrm{V_{Det}}$ \cite{Taubert2011}. The values for $\mathrm{V_{Bias}}$ and $\mathrm{V_{Det}}$ along a threshold current then yield the energy calibration of the detection barrier.

%\section{End Methods}

\begin{figure}
\includegraphics{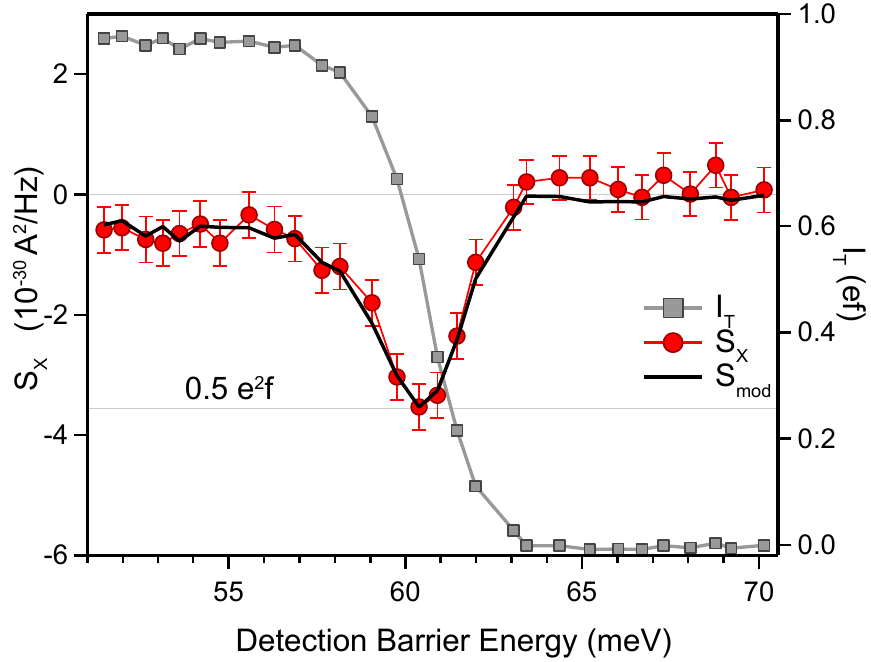}
\caption{\label{fig:fig2}
%\textbf{Partition noise of an on-demand electron source}
Transmitted current $\mathrm{I_T}$ and cross-correlation noise power $\mathrm{S_X}$ as a function of the energetic detector barrier height with the electron source configured to emit a sequence of single electrons. The error bars consist of the statistical error, the error of the estimated background, and the calibration error. The solid black line represents the expected value $\mathrm{S_{mod}}$ of the partitioning noise determined by the transmission probability $T(E)$,  which in turn is given by the fraction of the transmitted current.}
\end{figure}

The generation and energy-selective detection of on-demand non-equilibrium electrons is demonstrated with the electron source configured to emit one electron at a repetition frequency $f=\SI{280}{MHz}$. The transmitted current $\mathrm{I_T}$ as a function of the barrier energy is shown in Fig.~\ref{fig:fig2}. The maximum level of $\mathrm{I_T}$ is \SI{2}{\percent} below $1~ef$ due to residual inelastic scattering events on the \SI{2}{\micro m} long path to the barrier.
For energies greater than \SI{57}{meV} the current is pinched off as all of the emitted electrons are reflected at the beam splitter. The emission energy is defined by the exit barrier height and can be additionally controlled by amplitude and shape of the driving waveform, which also influences the width of the energy distribution \cite{Leicht2011}. 

Figure~\ref{fig:fig2} also shows the cross-correlation noise power $\mathrm{S_X}$ (red circles). As the sequence of emitted electrons is very regular and does not fluctuate in comparison to the resolution limit of our noise setup, the correlation between the two detectors is solely determined by the partitioning of the electrons. For energies below the pinch-off region we observe a small background of partitioning noise due to the \SI{2}{\%} reflection of the  energy-relaxed electrons at the detection barrier. When all electrons are reflected at large detector barrier energies, the cross-correlation signal is zero.

For one-electron scattering the current can be equated with the transmission probability $T(E)$ as a function of the electron energy $\mathrm{I_T}(E)=T(E)~ef $
and the resulting expected value for the partitioning noise $\mathrm{S}_{\mathrm{mod}} = -2~T(1-T)~e^2f$ (black line) agrees very well with the measured data. In particular, the negative sign of the cross-correlation signal reflects the anti-correlation of the two detector channels, as the sequence of single electrons is partitioned into either reflected or transmitted electrons.

\begin{figure*}
\includegraphics{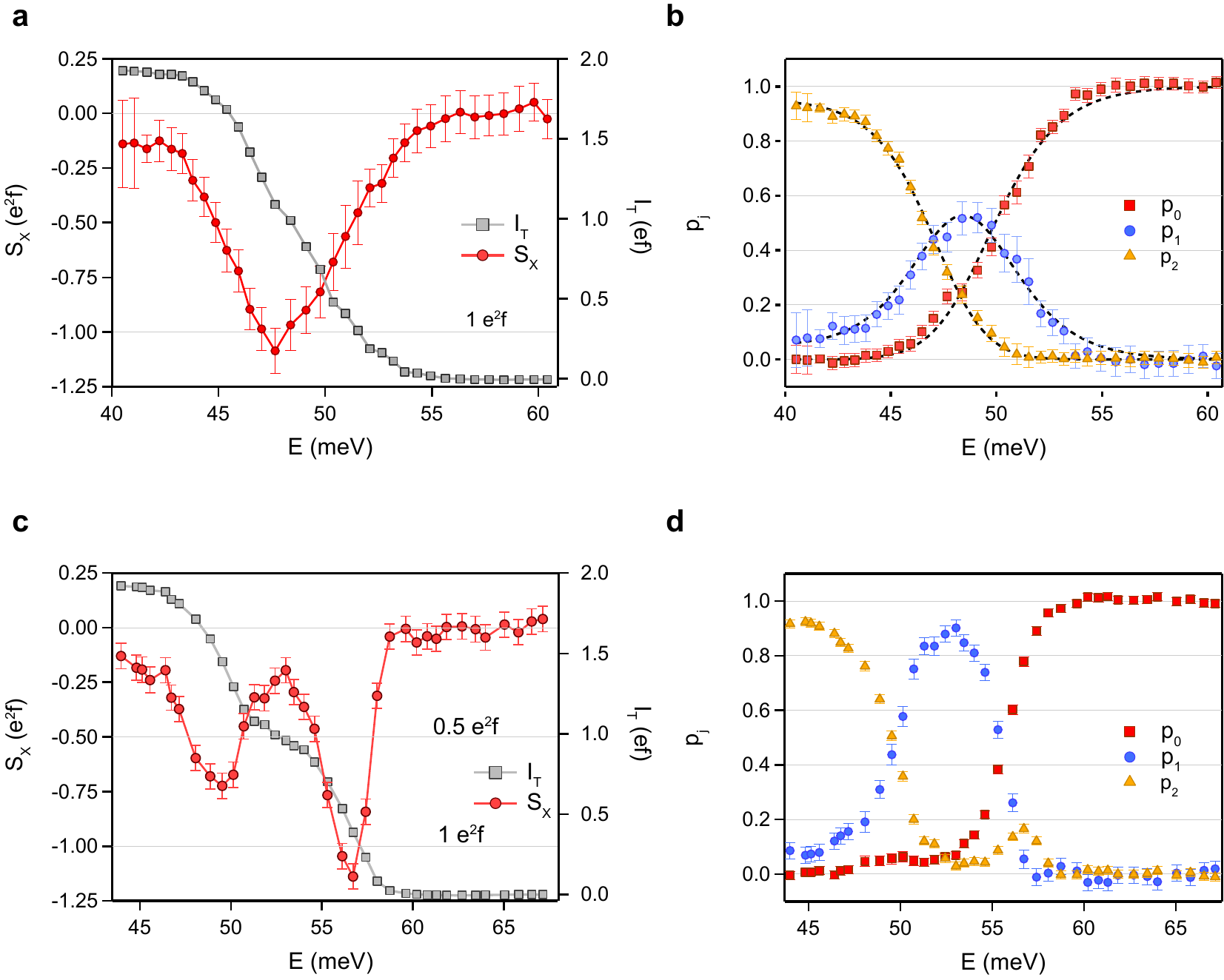}
\caption{\label{fig:fig3}
%\textbf{Partitioning of on-demand electron pairs}
\textbf{a} Transmitted current and cross-correlation noise power for a source of electron pairs. The horizontal line at $e^2f$ indicates the level for the superimposed noise amplitude of two partitioned electrons. \textbf{b} Probabilities for two transmitted (reflected) electrons $p_2$ ($p_0$) and for split electron pairs $p_1$. The dashed line represents a fit of the data with a binomial distribution. \textbf{c} Electron source tuned to emit electrons at separate energies leading to (\textbf{d}) a splitting fidelity of $90\%$ and an increased bunching at \SI{57}{meV} ($p_2$) and \SI{50}{meV} ($p_0$).}
\end{figure*}

In the following we characterize the on-demand emission and splitting of electron pairs. With the quantum dot current tuned on the $2~ef$ plateau one electron pair is emitted per waveform cycle. The corresponding partitioning results are  shown in Fig.~\ref{fig:fig3}a. The noise power shows a single dip with twice the amplitude compared to the single-electron case, cf. Fig.~\ref{fig:fig2}a.
For a clocked stream of electron pairs the full counting statistics of transmission can be unambiguously derived from the measurement of $\mathrm{I_T}$ and $\mathrm{S_X}$ : $\mathrm{I_T} = (p_1 + 2p_2)~ef$, $\mathrm{S_X} =  -2~[ p_0 (1-p_0) + p_2 (1-p_2) + 2 p_0 p_2 ]~e^2f $. $p_2$ and $p_0$ denote the probability that both electrons are transmitted or reflected. $p_1$ is the probability for the electron pair to split and both detector channels to record one electron each (Fig.~\ref{fig:fig3}b). If the two electrons are emitted in sequence at the same energy and subsequently scatter independently with the same probability $T$, the expected partitioning distribution is binomial, $p_2 = T^2$ and $p_0=(1-T)^2$. Assuming a simple step function $T(E)=1/\{1+\exp [( E-E_0)/\Delta_b] \}$ we derive the probabilities shown in Fig.~\ref{fig:fig3}b by a black dashed line, with $\Delta_b=\SI{1.8}{meV}$ as a fit parameter. A monotonic energy-dependence for the barrier yields good agreement with the measurement and at $T=0.5$ the maximum probability for splitting the electron pair is $p_1=50\%$.

The emission characteristics of the electron pair can be tuned by adjusting the driving oscillator waveform \cite{Giblin2012}. Specifically, we shorten the unloading phase with the goal of maintaining the energy difference in the emission spectrum imposed by the confinement in the quantum dot. As a result, the transmitted current shows a plateau at $1~ef$ hinting at a separation of the two electrons in energy (Fig.~\ref{fig:fig3}c). In comparison to the previous measurement the cross-correlation noise power now displays two separate dips at transmission current values $1.5~ef$ and $0.5~ef$. In between these dips the cross-correlation noise drops to a low absolute value of $0.2~e^2f$ at a transmission current of $1~ef$ consistent with each of the two detectors receiving a positively correlated, but highly regular stream of one electron per cycle. 
The corresponding counting statistics (Fig.~\ref{fig:fig3}d) shows the fidelity of splitting the electron pair ($p_1$) reaching \SI{90}{\%}.

The counting statistics not only allow  to verify the pair-splitting fidelity but can also provide evidence for electron interactions in the partitioning process.
Strong  anti-correlation between the paths taken by individual electrons after the pair  hits the barrier  ($p_1 >50~\%$) indicates that the two electrons are distinguished by the scattering process. Statistically independent scattering of two electrons with transmission probabilities $T_{a}$ and $T_b$
would result in a Poisson binomial distribution, $p_0=T_a T_b$, $p_2=(1-T_a)(1-T_b)$, which obeys an inequality $\sqrt{p_0}+\sqrt{p_2} \leq 1$. This necessary condition for statistical independence defines a domain in $(p_0, p_2)$ plane, which is compared to the statistics inferred from measurements in Fig.~4. The data for equally partitioned electrons (see Fig.~3b, blue circles in Fig.~4) fall on the border of the domain, as expected from $T_a=T_b=T$. The high-fidelity partitioning (Fig.~3d, red squares in Fig.~4) is  consistent with the Poisson binomial distribution for a semi-transparent barrier (small $p_0,p_2$), where the electrons are distinguished by $T_a \not =T_b$. 
Within this region an HBT correlation measurement is unable to discriminate the physical mechanisms of partitioning: a difference in transmission coefficients may indicate splitting in energy or a more complex orbital separation induced by the Pauli principle (supplementary note). 
However, for low transmittance (small $p_2$, large $p_0$) the independence condition is violated, indicating an increased bunching of electrons beyond statistical coincidence. This effect is reflected in the cross-correlation shot noise which significantly exceeds the value of single particle partitioning ($0.5~e^2f$) suggesting the two electrons are reflected or transmitted together at the respective energies ($\mathrm{E}=\SI{50}{meV}$ and $\mathrm{E}=\SI{57}{meV}$ in Fig.~3c and 3d).

The violation of scattering independence requires two-body correlations in the incoming state or in the partitioning process. Considering quantum correlations in the incoming state imposed by the Fermi statistics, we note that a single Slater determinant of two orthogonal spin-orbitals is insufficient for the counting statistics to exceed the binomial constraint while a spin-singlet with additional orbital correlations (e.g., a symmetric combination of two orbitals with different energies \cite{Hassler2007, Hassler2008}) can in principle cause a bunching effect in the transmission. However,
the non-monotonic energy dependence of the bunching anomaly (visible as a peak in $p_2(E)$ in Fig.~3d near $E =\SI{57}{meV}$) is inconsistent with the single-channel Landauer-B\"{u}ttiker picture and a monotonic $T(E)$ regardless of the initial correlations in the incoming two-electron state including an arbitrary mixed state. (supplementary note). 

We therefore conclude that electron-electron interaction in the partitioning process  causes the observed bunching, which opens a path to introduce nonlinearity in electron quantum optics devices. 

We thank H. Marx, Th. Weimann and P. Mirovsky for the fabrication of the wafer material and the device. The authors (except V.K.) acknowledge financial support by the Deutsche Forschungsgemeinschaft. V.K. has been supported by the Latvian Science Council.

%\nocite{*}

\bibliography{lit}

\begin{thebibliography}{32}%
\makeatletter
\providecommand \@ifxundefined [1]{%
 \@ifx{#1\undefined}
}%
\providecommand \@ifnum [1]{%
 \ifnum #1\expandafter \@firstoftwo
 \else \expandafter \@secondoftwo
 \fi
}%
\providecommand \@ifx [1]{%
 \ifx #1\expandafter \@firstoftwo
 \else \expandafter \@secondoftwo
 \fi
}%
\providecommand \natexlab [1]{#1}%
\providecommand \enquote  [1]{``#1''}%
\providecommand \bibnamefont  [1]{#1}%
\providecommand \bibfnamefont [1]{#1}%
\providecommand \citenamefont [1]{#1}%
\providecommand \href@noop [0]{\@secondoftwo}%
\providecommand \href [0]{\begingroup \@sanitize@url \@href}%
\providecommand \@href[1]{\@@startlink{#1}\@@href}%
\providecommand \@@href[1]{\endgroup#1\@@endlink}%
\providecommand \@sanitize@url [0]{\catcode `\\12\catcode `\$12\catcode
  `\&12\catcode `\#12\catcode `\^12\catcode `\_12\catcode `\%12\relax}%
\providecommand \@@startlink[1]{}%
\providecommand \@@endlink[0]{}%
\providecommand \url  [0]{\begingroup\@sanitize@url \@url }%
\providecommand \@url [1]{\endgroup\@href {#1}{\urlprefix }}%
\providecommand \urlprefix  [0]{URL }%
\providecommand \Eprint [0]{\href }%
\providecommand \doibase [0]{http://dx.doi.org/}%
\providecommand \selectlanguage [0]{\@gobble}%
\providecommand \bibinfo  [0]{\@secondoftwo}%
\providecommand \bibfield  [0]{\@secondoftwo}%
\providecommand \translation [1]{[#1]}%
\providecommand \BibitemOpen [0]{}%
\providecommand \bibitemStop [0]{}%
\providecommand \bibitemNoStop [0]{.\EOS\space}%
\providecommand \EOS [0]{\spacefactor3000\relax}%
\providecommand \BibitemShut  [1]{\csname bibitem#1\endcsname}%
\let\auto@bib@innerbib\@empty
%</preamble>
\bibitem [{\citenamefont {Kim}\ \emph {et~al.}(1999)\citenamefont {Kim},
  \citenamefont {Benson}, \citenamefont {Kan},\ and\ \citenamefont
  {Yamamoto}}]{Kim1999}%
  \BibitemOpen
  \bibfield  {author} {\bibinfo {author} {\bibfnamefont {J.}~\bibnamefont
  {Kim}}, \bibinfo {author} {\bibfnamefont {O.}~\bibnamefont {Benson}},
  \bibinfo {author} {\bibfnamefont {H.}~\bibnamefont {Kan}}, \ and\ \bibinfo
  {author} {\bibfnamefont {Y.}~\bibnamefont {Yamamoto}},\ }\href
  {http://dx.doi.org/10.1038/17295} {\bibfield  {journal} {\bibinfo  {journal}
  {Nature}\ }\textbf {\bibinfo {volume} {397}},\ \bibinfo {pages} {500}
  (\bibinfo {year} {1999})}\BibitemShut {NoStop}%
\bibitem [{\citenamefont {Stevenson}\ \emph {et~al.}(2006)\citenamefont
  {Stevenson}, \citenamefont {Young}, \citenamefont {Atkinson}, \citenamefont
  {Cooper}, \citenamefont {Ritchie},\ and\ \citenamefont
  {Shields}}]{Stevenson2006}%
  \BibitemOpen
  \bibfield  {author} {\bibinfo {author} {\bibfnamefont {R.~M.}\ \bibnamefont
  {Stevenson}}, \bibinfo {author} {\bibfnamefont {R.~J.}\ \bibnamefont
  {Young}}, \bibinfo {author} {\bibfnamefont {P.}~\bibnamefont {Atkinson}},
  \bibinfo {author} {\bibfnamefont {K.}~\bibnamefont {Cooper}}, \bibinfo
  {author} {\bibfnamefont {D.~A.}\ \bibnamefont {Ritchie}}, \ and\ \bibinfo
  {author} {\bibfnamefont {A.~J.}\ \bibnamefont {Shields}},\ }\href {\doibase
  10.1038/nature04446} {\bibfield  {journal} {\bibinfo  {journal} {Nature}\
  }\textbf {\bibinfo {volume} {439}},\ \bibinfo {pages} {179} (\bibinfo {year}
  {2006})}\BibitemShut {NoStop}%
\bibitem [{\citenamefont {Lang}\ \emph {et~al.}(2013)\citenamefont {Lang},
  \citenamefont {Eichler}, \citenamefont {Steffen}, \citenamefont {Fink},
  \citenamefont {Woolley}, \citenamefont {Blais},\ and\ \citenamefont
  {Wallraff}}]{Lang2013}%
  \BibitemOpen
  \bibfield  {author} {\bibinfo {author} {\bibfnamefont {C.}~\bibnamefont
  {Lang}}, \bibinfo {author} {\bibfnamefont {C.}~\bibnamefont {Eichler}},
  \bibinfo {author} {\bibfnamefont {L.}~\bibnamefont {Steffen}}, \bibinfo
  {author} {\bibfnamefont {J.~M.}\ \bibnamefont {Fink}}, \bibinfo {author}
  {\bibfnamefont {M.~J.}\ \bibnamefont {Woolley}}, \bibinfo {author}
  {\bibfnamefont {A.}~\bibnamefont {Blais}}, \ and\ \bibinfo {author}
  {\bibfnamefont {A.}~\bibnamefont {Wallraff}},\ }\href {\doibase
  10.1038/nphys2612} {\bibfield  {journal} {\bibinfo  {journal} {Nature
  Physics}\ }\textbf {\bibinfo {volume} {9}},\ \bibinfo {pages} {345} (\bibinfo
  {year} {2013})}\BibitemShut {NoStop}%
\bibitem [{\citenamefont {Yamamoto}\ \emph {et~al.}(2012)\citenamefont
  {Yamamoto}, \citenamefont {Takada}, \citenamefont {B\"auerle}, \citenamefont
  {Watanabe}, \citenamefont {Wieck},\ and\ \citenamefont
  {Tarucha}}]{Yamamoto2012}%
  \BibitemOpen
  \bibfield  {author} {\bibinfo {author} {\bibfnamefont {M.}~\bibnamefont
  {Yamamoto}}, \bibinfo {author} {\bibfnamefont {S.}~\bibnamefont {Takada}},
  \bibinfo {author} {\bibfnamefont {C.}~\bibnamefont {B\"auerle}}, \bibinfo
  {author} {\bibfnamefont {K.}~\bibnamefont {Watanabe}}, \bibinfo {author}
  {\bibfnamefont {A.~D.}\ \bibnamefont {Wieck}}, \ and\ \bibinfo {author}
  {\bibfnamefont {S.}~\bibnamefont {Tarucha}},\ }\href {\doibase
  10.1038/nnano.2012.28} {\bibfield  {journal} {\bibinfo  {journal} {Nature
  Nanotechnology}\ }\textbf {\bibinfo {volume} {7}},\ \bibinfo {pages} {247}
  (\bibinfo {year} {2012})}\BibitemShut {NoStop}%
\bibitem [{\citenamefont {Ji}\ \emph {et~al.}(2003)\citenamefont {Ji},
  \citenamefont {Chung}, \citenamefont {Sprinzak}, \citenamefont {Heiblum},
  \citenamefont {Mahalu},\ and\ \citenamefont {Shtrikman}}]{Ji2003}%
  \BibitemOpen
  \bibfield  {author} {\bibinfo {author} {\bibfnamefont {Y.}~\bibnamefont
  {Ji}}, \bibinfo {author} {\bibfnamefont {Y.}~\bibnamefont {Chung}}, \bibinfo
  {author} {\bibfnamefont {D.}~\bibnamefont {Sprinzak}}, \bibinfo {author}
  {\bibfnamefont {M.}~\bibnamefont {Heiblum}}, \bibinfo {author} {\bibfnamefont
  {D.}~\bibnamefont {Mahalu}}, \ and\ \bibinfo {author} {\bibfnamefont
  {H.}~\bibnamefont {Shtrikman}},\ }\href
  {http://dx.doi.org/10.1038/nature01503} {\bibfield  {journal} {\bibinfo
  {journal} {Nature}\ }\textbf {\bibinfo {volume} {422}},\ \bibinfo {pages}
  {415} (\bibinfo {year} {2003})}\BibitemShut {NoStop}%
\bibitem [{\citenamefont {Neder}\ \emph {et~al.}(2007)\citenamefont {Neder},
  \citenamefont {Ofek}, \citenamefont {Chung}, \citenamefont {Heiblum},
  \citenamefont {Mahalu},\ and\ \citenamefont {Umansky}}]{Neder2007}%
  \BibitemOpen
  \bibfield  {author} {\bibinfo {author} {\bibfnamefont {I.}~\bibnamefont
  {Neder}}, \bibinfo {author} {\bibfnamefont {N.}~\bibnamefont {Ofek}},
  \bibinfo {author} {\bibfnamefont {Y.}~\bibnamefont {Chung}}, \bibinfo
  {author} {\bibfnamefont {M.}~\bibnamefont {Heiblum}}, \bibinfo {author}
  {\bibfnamefont {D.}~\bibnamefont {Mahalu}}, \ and\ \bibinfo {author}
  {\bibfnamefont {V.}~\bibnamefont {Umansky}},\ }\href
  {http://dx.doi.org/10.1038/nature05955} {\bibfield  {journal} {\bibinfo
  {journal} {Nature}\ }\textbf {\bibinfo {volume} {448}},\ \bibinfo {pages}
  {333} (\bibinfo {year} {2007})}\BibitemShut {NoStop}%
\bibitem [{\citenamefont {Altimiras}\ \emph {et~al.}(2009)\citenamefont
  {Altimiras}, \citenamefont {le~Sueur}, \citenamefont {Gennser}, \citenamefont
  {Cavanna}, \citenamefont {Mailly},\ and\ \citenamefont
  {Pierre}}]{Altimiras2009}%
  \BibitemOpen
  \bibfield  {author} {\bibinfo {author} {\bibfnamefont {C.}~\bibnamefont
  {Altimiras}}, \bibinfo {author} {\bibfnamefont {H.}~\bibnamefont {le~Sueur}},
  \bibinfo {author} {\bibfnamefont {U.}~\bibnamefont {Gennser}}, \bibinfo
  {author} {\bibfnamefont {A.}~\bibnamefont {Cavanna}}, \bibinfo {author}
  {\bibfnamefont {D.}~\bibnamefont {Mailly}}, \ and\ \bibinfo {author}
  {\bibfnamefont {F.}~\bibnamefont {Pierre}},\ }\href {\doibase
  10.1038/nphys1429} {\bibfield  {journal} {\bibinfo  {journal} {Nature
  Physics}\ }\textbf {\bibinfo {volume} {6}},\ \bibinfo {pages} {34} (\bibinfo
  {year} {2009})}\BibitemShut {NoStop}%
\bibitem [{\citenamefont {Henny}\ \emph {et~al.}(1999)\citenamefont {Henny},
  \citenamefont {Oberholzer}, \citenamefont {Strunk}, \citenamefont {Heinzel},
  \citenamefont {Ensslin}, \citenamefont {Holland},\ and\ \citenamefont
  {Schonenberger}}]{Henny1999}%
  \BibitemOpen
  \bibfield  {author} {\bibinfo {author} {\bibfnamefont {M.}~\bibnamefont
  {Henny}}, \bibinfo {author} {\bibfnamefont {S.}~\bibnamefont {Oberholzer}},
  \bibinfo {author} {\bibfnamefont {C.}~\bibnamefont {Strunk}}, \bibinfo
  {author} {\bibfnamefont {T.}~\bibnamefont {Heinzel}}, \bibinfo {author}
  {\bibfnamefont {K.}~\bibnamefont {Ensslin}}, \bibinfo {author} {\bibfnamefont
  {M.}~\bibnamefont {Holland}}, \ and\ \bibinfo {author} {\bibfnamefont
  {C.}~\bibnamefont {Schonenberger}},\ }\href {\doibase
  10.1126/science.284.5412.296} {\bibfield  {journal} {\bibinfo  {journal}
  {Science}\ }\textbf {\bibinfo {volume} {284}},\ \bibinfo {pages} {296}
  (\bibinfo {year} {1999})}\BibitemShut {NoStop}%
\bibitem [{\citenamefont {Oliver}\ \emph {et~al.}(1999)\citenamefont {Oliver},
  \citenamefont {Kim}, \citenamefont {Liu},\ and\ \citenamefont
  {Yamamoto}}]{Oliver1999}%
  \BibitemOpen
  \bibfield  {author} {\bibinfo {author} {\bibfnamefont {W.~D.}\ \bibnamefont
  {Oliver}}, \bibinfo {author} {\bibfnamefont {J.}~\bibnamefont {Kim}},
  \bibinfo {author} {\bibfnamefont {R.~C.}\ \bibnamefont {Liu}}, \ and\
  \bibinfo {author} {\bibfnamefont {Y.}~\bibnamefont {Yamamoto}},\ }\href
  {\doibase 10.1126/science.284.5412.299} {\bibfield  {journal} {\bibinfo
  {journal} {Science}\ }\textbf {\bibinfo {volume} {284}},\ \bibinfo {pages}
  {299} (\bibinfo {year} {1999})}\BibitemShut {NoStop}%
\bibitem [{\citenamefont {Fève}\ \emph {et~al.}(2007)\citenamefont {Fève},
  \citenamefont {Mahé}, \citenamefont {Berroir}, \citenamefont {Kontos},
  \citenamefont {Plaçais}, \citenamefont {Glattli}, \citenamefont {Cavanna},
  \citenamefont {Etienne},\ and\ \citenamefont {Jin}}]{Feve2007}%
  \BibitemOpen
  \bibfield  {author} {\bibinfo {author} {\bibfnamefont {G.}~\bibnamefont
  {Fève}}, \bibinfo {author} {\bibfnamefont {A.}~\bibnamefont {Mahé}},
  \bibinfo {author} {\bibfnamefont {J.-M.}\ \bibnamefont {Berroir}}, \bibinfo
  {author} {\bibfnamefont {T.}~\bibnamefont {Kontos}}, \bibinfo {author}
  {\bibfnamefont {B.}~\bibnamefont {Plaçais}}, \bibinfo {author}
  {\bibfnamefont {D.~C.}\ \bibnamefont {Glattli}}, \bibinfo {author}
  {\bibfnamefont {A.}~\bibnamefont {Cavanna}}, \bibinfo {author} {\bibfnamefont
  {B.}~\bibnamefont {Etienne}}, \ and\ \bibinfo {author} {\bibfnamefont
  {Y.}~\bibnamefont {Jin}},\ }\href {\doibase 10.1126/science.1141243}
  {\bibfield  {journal} {\bibinfo  {journal} {Science}\ }\textbf {\bibinfo
  {volume} {316}},\ \bibinfo {pages} {1169} (\bibinfo {year}
  {2007})}\BibitemShut {NoStop}%
\bibitem [{\citenamefont {Bocquillon}\ \emph {et~al.}(2012)\citenamefont
  {Bocquillon}, \citenamefont {Parmentier}, \citenamefont {Grenier},
  \citenamefont {Berroir}, \citenamefont {Degiovanni}, \citenamefont {Glattli},
  \citenamefont {Pla\ifmmode~\mbox{\c{c}}\else \c{c}\fi{}ais}, \citenamefont
  {Cavanna}, \citenamefont {Jin},\ and\ \citenamefont
  {F\`eve}}]{Bocquillon2012}%
  \BibitemOpen
  \bibfield  {author} {\bibinfo {author} {\bibfnamefont {E.}~\bibnamefont
  {Bocquillon}}, \bibinfo {author} {\bibfnamefont {F.~D.}\ \bibnamefont
  {Parmentier}}, \bibinfo {author} {\bibfnamefont {C.}~\bibnamefont {Grenier}},
  \bibinfo {author} {\bibfnamefont {J.-M.}\ \bibnamefont {Berroir}}, \bibinfo
  {author} {\bibfnamefont {P.}~\bibnamefont {Degiovanni}}, \bibinfo {author}
  {\bibfnamefont {D.~C.}\ \bibnamefont {Glattli}}, \bibinfo {author}
  {\bibfnamefont {B.}~\bibnamefont {Pla\ifmmode~\mbox{\c{c}}\else
  \c{c}\fi{}ais}}, \bibinfo {author} {\bibfnamefont {A.}~\bibnamefont
  {Cavanna}}, \bibinfo {author} {\bibfnamefont {Y.}~\bibnamefont {Jin}}, \ and\
  \bibinfo {author} {\bibfnamefont {G.}~\bibnamefont {F\`eve}},\ }\href
  {\doibase 10.1103/PhysRevLett.108.196803} {\bibfield  {journal} {\bibinfo
  {journal} {Phys. Rev. Lett.}\ }\textbf {\bibinfo {volume} {108}},\ \bibinfo
  {pages} {196803} (\bibinfo {year} {2012})}\BibitemShut {NoStop}%
\bibitem [{\citenamefont {Fletcher}\ \emph {et~al.}(2013)\citenamefont
  {Fletcher}, \citenamefont {See}, \citenamefont {Howe}, \citenamefont
  {Pepper}, \citenamefont {Giblin}, \citenamefont {Griffiths}, \citenamefont
  {Jones}, \citenamefont {Farrer}, \citenamefont {Ritchie}, \citenamefont
  {Janssen},\ and\ \citenamefont {Kataoka}}]{Fletcher2013}%
  \BibitemOpen
  \bibfield  {author} {\bibinfo {author} {\bibfnamefont {J.~D.}\ \bibnamefont
  {Fletcher}}, \bibinfo {author} {\bibfnamefont {P.}~\bibnamefont {See}},
  \bibinfo {author} {\bibfnamefont {H.}~\bibnamefont {Howe}}, \bibinfo {author}
  {\bibfnamefont {M.}~\bibnamefont {Pepper}}, \bibinfo {author} {\bibfnamefont
  {S.~P.}\ \bibnamefont {Giblin}}, \bibinfo {author} {\bibfnamefont {J.~P.}\
  \bibnamefont {Griffiths}}, \bibinfo {author} {\bibfnamefont {G.~A.~C.}\
  \bibnamefont {Jones}}, \bibinfo {author} {\bibfnamefont {I.}~\bibnamefont
  {Farrer}}, \bibinfo {author} {\bibfnamefont {D.~A.}\ \bibnamefont {Ritchie}},
  \bibinfo {author} {\bibfnamefont {T.~J. B.~M.}\ \bibnamefont {Janssen}}, \
  and\ \bibinfo {author} {\bibfnamefont {M.}~\bibnamefont {Kataoka}},\ }\href
  {\doibase 10.1103/PhysRevLett.111.216807} {\bibfield  {journal} {\bibinfo
  {journal} {Phys. Rev. Lett.}\ }\textbf {\bibinfo {volume} {111}},\ \bibinfo
  {pages} {216807} (\bibinfo {year} {2013})}\BibitemShut {NoStop}%
\bibitem [{\citenamefont {Liu}\ \emph {et~al.}(1998)\citenamefont {Liu},
  \citenamefont {Odom}, \citenamefont {Yamamoto},\ and\ \citenamefont
  {Tarucha}}]{Liu1998}%
  \BibitemOpen
  \bibfield  {author} {\bibinfo {author} {\bibfnamefont {R.~C.}\ \bibnamefont
  {Liu}}, \bibinfo {author} {\bibfnamefont {B.}~\bibnamefont {Odom}}, \bibinfo
  {author} {\bibfnamefont {Y.}~\bibnamefont {Yamamoto}}, \ and\ \bibinfo
  {author} {\bibfnamefont {S.}~\bibnamefont {Tarucha}},\ }\href {\doibase
  10.1038/34611} {\bibfield  {journal} {\bibinfo  {journal} {Nature}\ }\textbf
  {\bibinfo {volume} {391}},\ \bibinfo {pages} {263} (\bibinfo {year}
  {1998})}\BibitemShut {NoStop}%
\bibitem [{\citenamefont {Bocquillon}\ \emph {et~al.}(2013)\citenamefont
  {Bocquillon}, \citenamefont {Freulon}, \citenamefont {Berroir}, \citenamefont
  {Degiovanni}, \citenamefont {Placais}, \citenamefont {Cavanna}, \citenamefont
  {Jin},\ and\ \citenamefont {Feve}}]{Bocquillon2013}%
  \BibitemOpen
  \bibfield  {author} {\bibinfo {author} {\bibfnamefont {E.}~\bibnamefont
  {Bocquillon}}, \bibinfo {author} {\bibfnamefont {V.}~\bibnamefont {Freulon}},
  \bibinfo {author} {\bibfnamefont {J.-M.}\ \bibnamefont {Berroir}}, \bibinfo
  {author} {\bibfnamefont {P.}~\bibnamefont {Degiovanni}}, \bibinfo {author}
  {\bibfnamefont {B.}~\bibnamefont {Placais}}, \bibinfo {author} {\bibfnamefont
  {A.}~\bibnamefont {Cavanna}}, \bibinfo {author} {\bibfnamefont
  {Y.}~\bibnamefont {Jin}}, \ and\ \bibinfo {author} {\bibfnamefont
  {G.}~\bibnamefont {Feve}},\ }\href {\doibase 10.1126/science.1232572}
  {\bibfield  {journal} {\bibinfo  {journal} {Science}\ }\textbf {\bibinfo
  {volume} {339}},\ \bibinfo {pages} {1054} (\bibinfo {year}
  {2013})}\BibitemShut {NoStop}%
\bibitem [{\citenamefont {Wei}\ and\ \citenamefont
  {Chandrasekhar}(2010)}]{Wei2010}%
  \BibitemOpen
  \bibfield  {author} {\bibinfo {author} {\bibfnamefont {J.}~\bibnamefont
  {Wei}}\ and\ \bibinfo {author} {\bibfnamefont {V.}~\bibnamefont
  {Chandrasekhar}},\ }\href {\doibase 10.1038/nphys1669} {\bibfield  {journal}
  {\bibinfo  {journal} {Nature Physics}\ }\textbf {\bibinfo {volume} {6}},\
  \bibinfo {pages} {494} (\bibinfo {year} {2010})}\BibitemShut {NoStop}%
\bibitem [{\citenamefont {Herrmann}\ \emph {et~al.}(2010)\citenamefont
  {Herrmann}, \citenamefont {Portier}, \citenamefont {Roche}, \citenamefont
  {Yeyati}, \citenamefont {Kontos},\ and\ \citenamefont
  {Strunk}}]{Herrmann2010}%
  \BibitemOpen
  \bibfield  {author} {\bibinfo {author} {\bibfnamefont {L.~G.}\ \bibnamefont
  {Herrmann}}, \bibinfo {author} {\bibfnamefont {F.}~\bibnamefont {Portier}},
  \bibinfo {author} {\bibfnamefont {P.}~\bibnamefont {Roche}}, \bibinfo
  {author} {\bibfnamefont {A.~L.}\ \bibnamefont {Yeyati}}, \bibinfo {author}
  {\bibfnamefont {T.}~\bibnamefont {Kontos}}, \ and\ \bibinfo {author}
  {\bibfnamefont {C.}~\bibnamefont {Strunk}},\ }\href {\doibase
  10.1103/PhysRevLett.104.026801} {\bibfield  {journal} {\bibinfo  {journal}
  {Phys. Rev. Lett.}\ }\textbf {\bibinfo {volume} {104}},\ \bibinfo {pages}
  {026801} (\bibinfo {year} {2010})}\BibitemShut {NoStop}%
\bibitem [{\citenamefont {Hofstetter}\ \emph {et~al.}(2009)\citenamefont
  {Hofstetter}, \citenamefont {Csonka}, \citenamefont {Nygard},\ and\
  \citenamefont {Schonenberger}}]{Hofstetter2009}%
  \BibitemOpen
  \bibfield  {author} {\bibinfo {author} {\bibfnamefont {L.}~\bibnamefont
  {Hofstetter}}, \bibinfo {author} {\bibfnamefont {S.}~\bibnamefont {Csonka}},
  \bibinfo {author} {\bibfnamefont {J.}~\bibnamefont {Nygard}}, \ and\ \bibinfo
  {author} {\bibfnamefont {C.}~\bibnamefont {Schonenberger}},\ }\href
  {http://dx.doi.org/10.1038/nature08432} {\bibfield  {journal} {\bibinfo
  {journal} {Nature}\ }\textbf {\bibinfo {volume} {461}},\ \bibinfo {pages}
  {960} (\bibinfo {year} {2009})}\BibitemShut {NoStop}%
\bibitem [{\citenamefont {Hofstetter}\ \emph {et~al.}(2011)\citenamefont
  {Hofstetter}, \citenamefont {Csonka}, \citenamefont {Baumgartner},
  \citenamefont {F\"ul\"op}, \citenamefont {d'Hollosy}, \citenamefont
  {Nyg\aa{}rd},\ and\ \citenamefont {Sch\"onenberger}}]{Hofstetter2011}%
  \BibitemOpen
  \bibfield  {author} {\bibinfo {author} {\bibfnamefont {L.}~\bibnamefont
  {Hofstetter}}, \bibinfo {author} {\bibfnamefont {S.}~\bibnamefont {Csonka}},
  \bibinfo {author} {\bibfnamefont {A.}~\bibnamefont {Baumgartner}}, \bibinfo
  {author} {\bibfnamefont {G.}~\bibnamefont {F\"ul\"op}}, \bibinfo {author}
  {\bibfnamefont {S.}~\bibnamefont {d'Hollosy}}, \bibinfo {author}
  {\bibfnamefont {J.}~\bibnamefont {Nyg\aa{}rd}}, \ and\ \bibinfo {author}
  {\bibfnamefont {C.}~\bibnamefont {Sch\"onenberger}},\ }\href {\doibase
  10.1103/PhysRevLett.107.136801} {\bibfield  {journal} {\bibinfo  {journal}
  {Phys. Rev. Lett.}\ }\textbf {\bibinfo {volume} {107}},\ \bibinfo {pages}
  {136801} (\bibinfo {year} {2011})}\BibitemShut {NoStop}%
\bibitem [{\citenamefont {Das}\ \emph {et~al.}(2012)\citenamefont {Das},
  \citenamefont {Ronen}, \citenamefont {Heiblum}, \citenamefont {Mahalu},
  \citenamefont {Kretinin},\ and\ \citenamefont {Shtrikman}}]{Das2012}%
  \BibitemOpen
  \bibfield  {author} {\bibinfo {author} {\bibfnamefont {A.}~\bibnamefont
  {Das}}, \bibinfo {author} {\bibfnamefont {Y.}~\bibnamefont {Ronen}}, \bibinfo
  {author} {\bibfnamefont {M.}~\bibnamefont {Heiblum}}, \bibinfo {author}
  {\bibfnamefont {D.}~\bibnamefont {Mahalu}}, \bibinfo {author} {\bibfnamefont
  {A.~V.}\ \bibnamefont {Kretinin}}, \ and\ \bibinfo {author} {\bibfnamefont
  {H.}~\bibnamefont {Shtrikman}},\ }\href {\doibase 10.1038/ncomms2169}
  {\bibfield  {journal} {\bibinfo  {journal} {Nature Communications}\ }\textbf
  {\bibinfo {volume} {3}},\ \bibinfo {pages} {1165} (\bibinfo {year}
  {2012})}\BibitemShut {NoStop}%
\bibitem [{\citenamefont {Schindele}\ \emph {et~al.}(2012)\citenamefont
  {Schindele}, \citenamefont {Baumgartner},\ and\ \citenamefont
  {Sch\"onenberger}}]{Schindele2012}%
  \BibitemOpen
  \bibfield  {author} {\bibinfo {author} {\bibfnamefont {J.}~\bibnamefont
  {Schindele}}, \bibinfo {author} {\bibfnamefont {A.}~\bibnamefont
  {Baumgartner}}, \ and\ \bibinfo {author} {\bibfnamefont {C.}~\bibnamefont
  {Sch\"onenberger}},\ }\href {\doibase 10.1103/PhysRevLett.109.157002}
  {\bibfield  {journal} {\bibinfo  {journal} {Phys. Rev. Lett.}\ }\textbf
  {\bibinfo {volume} {109}},\ \bibinfo {pages} {157002} (\bibinfo {year}
  {2012})}\BibitemShut {NoStop}%
\bibitem [{\citenamefont {Burkard}\ \emph {et~al.}(2000)\citenamefont
  {Burkard}, \citenamefont {Loss},\ and\ \citenamefont
  {Sukhorukov}}]{Burkard2000}%
  \BibitemOpen
  \bibfield  {author} {\bibinfo {author} {\bibfnamefont {G.}~\bibnamefont
  {Burkard}}, \bibinfo {author} {\bibfnamefont {D.}~\bibnamefont {Loss}}, \
  and\ \bibinfo {author} {\bibfnamefont {E.~V.}\ \bibnamefont {Sukhorukov}},\
  }\href {\doibase 10.1103/PhysRevB.61.R16303} {\bibfield  {journal} {\bibinfo
  {journal} {Phys. Rev. B}\ }\textbf {\bibinfo {volume} {61}},\ \bibinfo
  {pages} {R16303} (\bibinfo {year} {2000})}\BibitemShut {NoStop}%
\bibitem [{\citenamefont {Samuelsson}\ and\ \citenamefont
  {B\"uttiker}(2006)}]{Samuelsson2006}%
  \BibitemOpen
  \bibfield  {author} {\bibinfo {author} {\bibfnamefont {P.}~\bibnamefont
  {Samuelsson}}\ and\ \bibinfo {author} {\bibfnamefont {M.}~\bibnamefont
  {B\"uttiker}},\ }\href {\doibase 10.1103/PhysRevB.73.041305} {\bibfield
  {journal} {\bibinfo  {journal} {Phys. Rev. B}\ }\textbf {\bibinfo {volume}
  {73}},\ \bibinfo {pages} {041305} (\bibinfo {year} {2006})}\BibitemShut
  {NoStop}%
\bibitem [{\citenamefont {Hassler}\ \emph {et~al.}(2007)\citenamefont
  {Hassler}, \citenamefont {Lesovik},\ and\ \citenamefont
  {Blatter}}]{Hassler2007}%
  \BibitemOpen
  \bibfield  {author} {\bibinfo {author} {\bibfnamefont {F.}~\bibnamefont
  {Hassler}}, \bibinfo {author} {\bibfnamefont {G.~B.}\ \bibnamefont
  {Lesovik}}, \ and\ \bibinfo {author} {\bibfnamefont {G.}~\bibnamefont
  {Blatter}},\ }\href {\doibase 10.1103/PhysRevLett.99.076804} {\bibfield
  {journal} {\bibinfo  {journal} {Phys. Rev. Lett.}\ }\textbf {\bibinfo
  {volume} {99}},\ \bibinfo {pages} {076804} (\bibinfo {year}
  {2007})}\BibitemShut {NoStop}%
\bibitem [{\citenamefont {Hassler}\ \emph {et~al.}(2008)\citenamefont
  {Hassler}, \citenamefont {Suslov}, \citenamefont {Graf}, \citenamefont
  {Lebedev}, \citenamefont {Lesovik},\ and\ \citenamefont
  {Blatter}}]{Hassler2008}%
  \BibitemOpen
  \bibfield  {author} {\bibinfo {author} {\bibfnamefont {F.}~\bibnamefont
  {Hassler}}, \bibinfo {author} {\bibfnamefont {M.~V.}\ \bibnamefont {Suslov}},
  \bibinfo {author} {\bibfnamefont {G.~M.}\ \bibnamefont {Graf}}, \bibinfo
  {author} {\bibfnamefont {M.~V.}\ \bibnamefont {Lebedev}}, \bibinfo {author}
  {\bibfnamefont {G.~B.}\ \bibnamefont {Lesovik}}, \ and\ \bibinfo {author}
  {\bibfnamefont {G.}~\bibnamefont {Blatter}},\ }\href {\doibase
  10.1103/PhysRevB.78.165330} {\bibfield  {journal} {\bibinfo  {journal} {Phys.
  Rev. B}\ }\textbf {\bibinfo {volume} {78}},\ \bibinfo {pages} {165330}
  (\bibinfo {year} {2008})}\BibitemShut {NoStop}%
\bibitem [{\citenamefont {Wahl}\ \emph {et~al.}(2014)\citenamefont {Wahl},
  \citenamefont {Rech}, \citenamefont {Jonckheere},\ and\ \citenamefont
  {Martin}}]{Wahl2014}%
  \BibitemOpen
  \bibfield  {author} {\bibinfo {author} {\bibfnamefont {C.}~\bibnamefont
  {Wahl}}, \bibinfo {author} {\bibfnamefont {J.}~\bibnamefont {Rech}}, \bibinfo
  {author} {\bibfnamefont {T.}~\bibnamefont {Jonckheere}}, \ and\ \bibinfo
  {author} {\bibfnamefont {T.}~\bibnamefont {Martin}},\ }\href {\doibase
  10.1103/PhysRevLett.112.046802} {\bibfield  {journal} {\bibinfo  {journal}
  {Phys. Rev. Lett.}\ }\textbf {\bibinfo {volume} {112}},\ \bibinfo {pages}
  {046802} (\bibinfo {year} {2014})}\BibitemShut {NoStop}%
\bibitem [{\citenamefont {Blumenthal}\ \emph {et~al.}(2007)\citenamefont
  {Blumenthal}, \citenamefont {Kaestner}, \citenamefont {Li}, \citenamefont
  {Giblin}, \citenamefont {Janssen}, \citenamefont {Pepper}, \citenamefont
  {Anderson}, \citenamefont {Jones},\ and\ \citenamefont
  {Ritchie}}]{Blumenthal2007}%
  \BibitemOpen
  \bibfield  {author} {\bibinfo {author} {\bibfnamefont {M.~D.}\ \bibnamefont
  {Blumenthal}}, \bibinfo {author} {\bibfnamefont {B.}~\bibnamefont
  {Kaestner}}, \bibinfo {author} {\bibfnamefont {L.}~\bibnamefont {Li}},
  \bibinfo {author} {\bibfnamefont {S.}~\bibnamefont {Giblin}}, \bibinfo
  {author} {\bibfnamefont {T.~J. B.~M.}\ \bibnamefont {Janssen}}, \bibinfo
  {author} {\bibfnamefont {M.}~\bibnamefont {Pepper}}, \bibinfo {author}
  {\bibfnamefont {D.}~\bibnamefont {Anderson}}, \bibinfo {author}
  {\bibfnamefont {G.}~\bibnamefont {Jones}}, \ and\ \bibinfo {author}
  {\bibfnamefont {D.~A.}\ \bibnamefont {Ritchie}},\ }\href {\doibase
  10.1038/nphys582} {\bibfield  {journal} {\bibinfo  {journal} {Nature
  Physics}\ }\textbf {\bibinfo {volume} {3}},\ \bibinfo {pages} {343} (\bibinfo
  {year} {2007})}\BibitemShut {NoStop}%
\bibitem [{\citenamefont {Kaestner}\ \emph {et~al.}(2008)\citenamefont
  {Kaestner}, \citenamefont {Kashcheyevs}, \citenamefont {Amakawa},
  \citenamefont {Blumenthal}, \citenamefont {Li}, \citenamefont {Janssen},
  \citenamefont {Hein}, \citenamefont {Pierz}, \citenamefont {Weimann},
  \citenamefont {Siegner},\ and\ \citenamefont {Schumacher}}]{Kaestner2008}%
  \BibitemOpen
  \bibfield  {author} {\bibinfo {author} {\bibfnamefont {B.}~\bibnamefont
  {Kaestner}}, \bibinfo {author} {\bibfnamefont {V.}~\bibnamefont
  {Kashcheyevs}}, \bibinfo {author} {\bibfnamefont {S.}~\bibnamefont
  {Amakawa}}, \bibinfo {author} {\bibfnamefont {M.~D.}\ \bibnamefont
  {Blumenthal}}, \bibinfo {author} {\bibfnamefont {L.}~\bibnamefont {Li}},
  \bibinfo {author} {\bibfnamefont {T.~J. B.~M.}\ \bibnamefont {Janssen}},
  \bibinfo {author} {\bibfnamefont {G.}~\bibnamefont {Hein}}, \bibinfo {author}
  {\bibfnamefont {K.}~\bibnamefont {Pierz}}, \bibinfo {author} {\bibfnamefont
  {T.}~\bibnamefont {Weimann}}, \bibinfo {author} {\bibfnamefont
  {U.}~\bibnamefont {Siegner}}, \ and\ \bibinfo {author} {\bibfnamefont
  {H.~W.}\ \bibnamefont {Schumacher}},\ }\href {\doibase
  10.1103/PhysRevB.77.153301} {\bibfield  {journal} {\bibinfo  {journal} {Phys.
  Rev. B}\ }\textbf {\bibinfo {volume} {77}},\ \bibinfo {eid} {153301}
  (\bibinfo {year} {2008})}\BibitemShut {NoStop}%
\bibitem [{\citenamefont {Giblin}\ \emph {et~al.}(2012)\citenamefont {Giblin},
  \citenamefont {Kataoka}, \citenamefont {Fletcher}, \citenamefont {See},
  \citenamefont {Janssen}, \citenamefont {Griffiths}, \citenamefont {Jones},
  \citenamefont {Farrer},\ and\ \citenamefont {Ritchie}}]{Giblin2012}%
  \BibitemOpen
  \bibfield  {author} {\bibinfo {author} {\bibfnamefont {S.~P.}\ \bibnamefont
  {Giblin}}, \bibinfo {author} {\bibfnamefont {M.}~\bibnamefont {Kataoka}},
  \bibinfo {author} {\bibfnamefont {J.~D.}\ \bibnamefont {Fletcher}}, \bibinfo
  {author} {\bibfnamefont {P.}~\bibnamefont {See}}, \bibinfo {author}
  {\bibfnamefont {T.~J. B.~M.}\ \bibnamefont {Janssen}}, \bibinfo {author}
  {\bibfnamefont {J.~P.}\ \bibnamefont {Griffiths}}, \bibinfo {author}
  {\bibfnamefont {G.~A.~C.}\ \bibnamefont {Jones}}, \bibinfo {author}
  {\bibfnamefont {I.}~\bibnamefont {Farrer}}, \ and\ \bibinfo {author}
  {\bibfnamefont {D.~A.}\ \bibnamefont {Ritchie}},\ }\href {\doibase
  10.1038/ncomms1935} {\bibfield  {journal} {\bibinfo  {journal} {Nature
  Communications}\ }\textbf {\bibinfo {volume} {3}},\ \bibinfo {pages} {930}
  (\bibinfo {year} {2012})}\BibitemShut {NoStop}%
\bibitem [{\citenamefont {Leicht}\ \emph {et~al.}(2011)\citenamefont {Leicht},
  \citenamefont {Mirovsky}, \citenamefont {Kaestner}, \citenamefont {Hohls},
  \citenamefont {Kashcheyevs}, \citenamefont {Kurganova}, \citenamefont
  {Zeitler}, \citenamefont {Weimann}, \citenamefont {Pierz},\ and\
  \citenamefont {Schumacher}}]{Leicht2011}%
  \BibitemOpen
  \bibfield  {author} {\bibinfo {author} {\bibfnamefont {C.}~\bibnamefont
  {Leicht}}, \bibinfo {author} {\bibfnamefont {P.}~\bibnamefont {Mirovsky}},
  \bibinfo {author} {\bibfnamefont {B.}~\bibnamefont {Kaestner}}, \bibinfo
  {author} {\bibfnamefont {F.}~\bibnamefont {Hohls}}, \bibinfo {author}
  {\bibfnamefont {V.}~\bibnamefont {Kashcheyevs}}, \bibinfo {author}
  {\bibfnamefont {E.~V.}\ \bibnamefont {Kurganova}}, \bibinfo {author}
  {\bibfnamefont {U.}~\bibnamefont {Zeitler}}, \bibinfo {author} {\bibfnamefont
  {T.}~\bibnamefont {Weimann}}, \bibinfo {author} {\bibfnamefont
  {K.}~\bibnamefont {Pierz}}, \ and\ \bibinfo {author} {\bibfnamefont {H.~W.}\
  \bibnamefont {Schumacher}},\ }\href {\doibase 10.1088/0268-1242/26/5/055010}
  {\bibfield  {journal} {\bibinfo  {journal} {Semiconductor Science and
  Technology}\ }\textbf {\bibinfo {volume} {26}},\ \bibinfo {pages} {055010}
  (\bibinfo {year} {2011})}\BibitemShut {NoStop}%
\bibitem [{\citenamefont {Maire}\ \emph {et~al.}(2008)\citenamefont {Maire},
  \citenamefont {Hohls}, \citenamefont {Kaestner}, \citenamefont {Pierz},
  \citenamefont {Schumacher},\ and\ \citenamefont {Haug}}]{Maire2008}%
  \BibitemOpen
  \bibfield  {author} {\bibinfo {author} {\bibfnamefont {N.}~\bibnamefont
  {Maire}}, \bibinfo {author} {\bibfnamefont {F.}~\bibnamefont {Hohls}},
  \bibinfo {author} {\bibfnamefont {B.}~\bibnamefont {Kaestner}}, \bibinfo
  {author} {\bibfnamefont {K.}~\bibnamefont {Pierz}}, \bibinfo {author}
  {\bibfnamefont {H.~W.}\ \bibnamefont {Schumacher}}, \ and\ \bibinfo {author}
  {\bibfnamefont {R.~J.}\ \bibnamefont {Haug}},\ }\href {\doibase
  10.1063/1.2885076} {\bibfield  {journal} {\bibinfo  {journal} {Applied
  Physics Letters}\ }\textbf {\bibinfo {volume} {92}},\ \bibinfo {eid} {082112}
  (\bibinfo {year} {2008})}\BibitemShut {NoStop}%
\bibitem [{\citenamefont {Taubert}\ \emph {et~al.}(2011)\citenamefont
  {Taubert}, \citenamefont {Tomaras}, \citenamefont {Schinner}, \citenamefont
  {Tranitz}, \citenamefont {Wegscheider}, \citenamefont {Kehrein},\ and\
  \citenamefont {Ludwig}}]{Taubert2011}%
  \BibitemOpen
  \bibfield  {author} {\bibinfo {author} {\bibfnamefont {D.}~\bibnamefont
  {Taubert}}, \bibinfo {author} {\bibfnamefont {C.}~\bibnamefont {Tomaras}},
  \bibinfo {author} {\bibfnamefont {G.~J.}\ \bibnamefont {Schinner}}, \bibinfo
  {author} {\bibfnamefont {H.~P.}\ \bibnamefont {Tranitz}}, \bibinfo {author}
  {\bibfnamefont {W.}~\bibnamefont {Wegscheider}}, \bibinfo {author}
  {\bibfnamefont {S.}~\bibnamefont {Kehrein}}, \ and\ \bibinfo {author}
  {\bibfnamefont {S.}~\bibnamefont {Ludwig}},\ }\href {\doibase
  10.1103/PhysRevB.83.235404} {\bibfield  {journal} {\bibinfo  {journal} {Phys.
  Rev. B}\ }\textbf {\bibinfo {volume} {83}},\ \bibinfo {pages} {235404}
  (\bibinfo {year} {2011})}\BibitemShut {NoStop}%
\bibitem [{\citenamefont {Sivan}\ \emph {et~al.}(1989)\citenamefont {Sivan},
  \citenamefont {Heiblum},\ and\ \citenamefont {Umbach}}]{Sivan1989}%
  \BibitemOpen
  \bibfield  {author} {\bibinfo {author} {\bibfnamefont {U.}~\bibnamefont
  {Sivan}}, \bibinfo {author} {\bibfnamefont {M.}~\bibnamefont {Heiblum}}, \
  and\ \bibinfo {author} {\bibfnamefont {C.~P.}\ \bibnamefont {Umbach}},\
  }\href {\doibase 10.1103/PhysRevLett.63.992} {\bibfield  {journal} {\bibinfo
  {journal} {Phys. Rev. Lett.}\ }\textbf {\bibinfo {volume} {63}},\ \bibinfo
  {pages} {992} (\bibinfo {year} {1989})}\BibitemShut {NoStop}%
\end{thebibliography}%

\newpage

\begin{figure}
\includegraphics{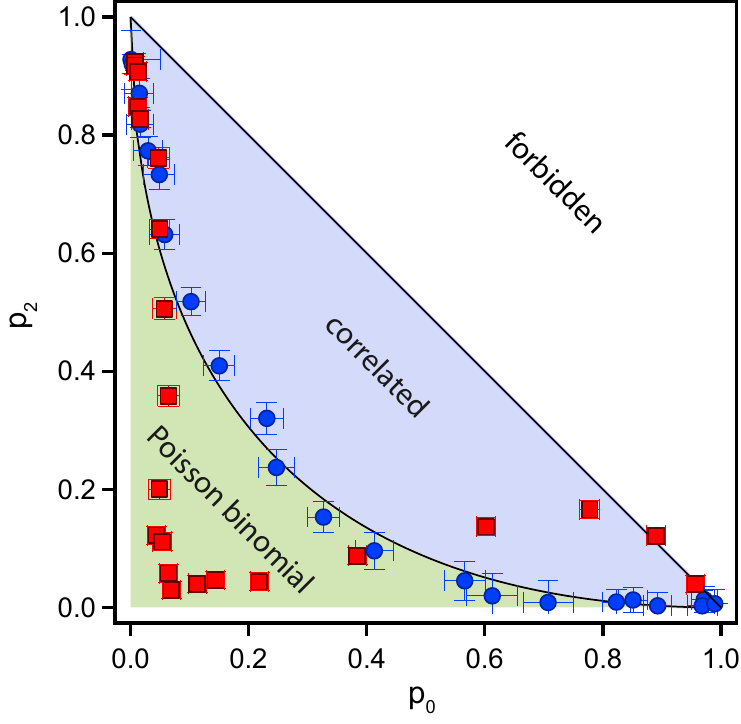}
\caption{\label{fig:fig4}
%\textbf{Regimes of binomial distribution}
$p_2$ as a function of $p_0$ for the two electron partitioning of Fig.~\ref{fig:fig3}b (blue circles) and Fig.~\ref{fig:fig3}d (red squares).  The line  $\sqrt{p_0}+\sqrt{p_2} = 1$ separates the domains of necessarily correlated and potentially independent partitioning.}
\end{figure}

\newcommand{\bra}[1]{\langle {#1} \rvert }

\renewcommand{\theequation}{S\arabic{equation}}%

\onecolumngrid
\appendix

\section*{Supplementary note}

%\section*{Counting statistics for single-channel scattering}
Here we assess the effect of initial correlations on the counting statistics of two electrons partitioned by an energy barrier which does not induce additional two-body correlations.
Specifically, we test the following combination of model assumptions:
(a) absence of two-body interactions at the barrier;
(b) single chiral orbital channel per spin projection $\sigma =\uparrow, \downarrow$, and (c) monotonic dependence of the single-electron scattering probability $T_{\sigma}(E)$ on the energy $E$ as tuned by the barrier-creating gate.
% Several previous theoretical investigations of the counting statistics in an HBT setup have followed the assumptions (a)-(c) \cite{Haasle2008}.

Assumption (a) implies that the quantum numbers that diagonalize the single-particle scattering matrix remain good
for the two-particle transmission. This allows one to express the two-particle scattering probabilities as 
\begin{equation}\label{eq:pindependent}
\begin{aligned}
  p_2 & =\sum\limits_{kq} \sum_{\sigma \sigma'} \rho_{k\sigma q\sigma'} T_{k\sigma} T_{q\sigma'} , \\
  p_0 & =\sum\limits_{kq} \sum_{\sigma \sigma'} \rho_{k\sigma q\sigma'} (1-T_{k\sigma})(1-T_{q\sigma'}) , 
\end{aligned} 
\end{equation}
where $0 \le T_{k\sigma} \leq 1$ are the transmission coefficients of the scattering eigenmodes and $\rho_{k\sigma q\sigma'}$ is the diagonal element of the two-particle density matrix of the incoming state. The third component of the statistics is $p_1 =1 -p_0-p_2$.
The necessary conditions on $\rho_{k\sigma q\sigma'}$
are non-negativity, normalization ($\sum_{k\sigma q\sigma'}  \rho_{k\sigma q\sigma'} =2$), and exchange symmetry ($\rho_{k\sigma q\sigma'}=
\rho_{ q\sigma'k\sigma}$).  (The latter is due to electrons being identical particles, not necessarily in identical states).

\subsection{Monotonicity argument}
Assumption (b) establishes that the orbital indices $k$ are not degenerate in energy. If $T_{k\sigma} =T_{\sigma}( \epsilon_{k\sigma} +E)$ where $\epsilon_{k\sigma}$ is the kinetic energy of the plane-wave mode with respect to a fixed reference level, and  $E$ is tuned linearly by the  barrier gate, then by assumption (c) and by the nonnegativity of $\rho$, the probabilities $p_2(E)$ and $p_0(E)$ have to be monotonic in $E$ as well.

\subsection{Effect of exchange correlations on the counting statistics}
For a pure state composed of two orthogonal spinorbitals,  $\ket{\Psi} = \frac{1}{\sqrt{2}} \left ( 
\ket{a}\otimes \ket{b}- \ket{b}\otimes \ket{a} \right )$ (a Slater determinant), one has 
%\begin{align}
 $ \rho_{k\sigma q\sigma'} = |\bra{\Psi} (\ket{k \sigma} \otimes\ket{q \sigma'} ) |^2 = \frac{1}{2} 
 | \qav{a| k\sigma} \qav{b|q\sigma'}- \qav{b|k\sigma}\qav{a|q\sigma'} |^2$
%\end{align} 
and
the counting statistics %\ref{eq:pindependent} 
(S1) become
\begin{align}
  p_2 = \det \begin{bmatrix}
               \qav{a | \hat{T} | a} & \qav{a | \hat{T} | b} \\
               \qav{b | \hat{T} | a} & \qav{b | \hat{T} | b}  \\
             \end{bmatrix}  , \quad 
  p_0 = \det \begin{bmatrix}
               1-\qav{a |\hat{T}| a} & \qav{a | \hat{T} | b} \\
               \qav{b | \hat{T} | a} & 1-\qav{b | \hat{T} | b}  \\
             \end{bmatrix}  ,         
\end{align}
where $\qav{a | \hat{T} | a}$ are matrix elements of a single-particle operator $\hat{T} =\sum_{k\sigma} \ket{k\sigma} T_{k\sigma} \bra{k\sigma}$.
Expressing the determinants in terms of products of eigenvalues,  we arrive at generalized binomial distribution $p_2=T_a T_b$ and $p_0=(1-T_a)(1-T_b)$ 
It easily follows from $0\leq p_i \leq 1$ that the eigenvalues satisfy $0 \leq T_a , T_b \leq 1$. For a derivation employing the generating function for counting statistics, see \cite{Hassler2007,Hassler2008}.

Eigenstates of the total spin operator that carry minimal entanglement 
are the spin singlet state with a doubly occupied orbital and the fully polarized two-orbital triplet states,
\begin{align}
 \ket{S} & \propto \left(\ket{\uparrow \downarrow}-\ket{\downarrow\uparrow}\right) \ket{\psi\psi} \, , \nonumber \\
 \ket{T_\sigma} & \propto \ket{\sigma\sigma} \left(\ket{\psi_1\psi_2}-\ket{\psi_2\psi_1}\right)  \, . \nonumber 
\end{align}
Since $\ket{S}$ and $\ket{T_\sigma}$ are expressible by a single Slater determinant,
their partitioning by an interactions-free barrier will also result 
in a generalized binomial distribution, regardless of the barrier-induced overlap $\qav{a | \hat{T} | b}$.
If the scattering is spin-independent, then the
same applies to the non-spin-polarized component of a two-orbital triplet, $\ket{T_0} \propto \left(\ket{\uparrow\downarrow}+\ket{\downarrow\uparrow}\right) \left(\ket{\psi_1\psi_2}-\ket{\psi_2\psi_1}\right)$.

Hence, only the states that carry more correlations than required by the entanglement due to exchange statistics 
may violate the constraint $\sqrt{p_0} +\sqrt{p_2} \leq 1$ under the single-particle scattering assumption (a).
An example of such a state (considered in detail in \cite{Hassler2007}) would be a singlet state with a non-factorizable exchange-symmetric orbital part.

\end{document}